%% file: paper.tex
\documentclass[10pt,conference]{IEEEtran}

\usepackage{tikz-cd}
\usepackage{tabularx}
\usepackage{listings, color}
\usepackage{latexsym}
\usepackage[T1]{fontenc}
\usepackage{balance}

\input{macros}
\input{macros_spl}

\begin{document}

\title{Towards Modal Software Engineering}

\author{\IEEEauthorblockN{Ramy Shahin}
		\IEEEauthorblockA{
			University of Toronto, Canada \\
			rshahin@cs.toronto.edu
		}
	}

\maketitle

\begin{abstract}
In this paper we introduce the notion of Modal Software Engineering: automatically turning sequential, deterministic programs into semantically equivalent programs efficiently operating on inputs coming from multiple overlapping worlds. We are drawing an analogy between modal logics, and software application domains where multiple sets of inputs (multiple worlds) need to be processed efficiently. Typically those sets highly overlap, so processing them independently would involve a lot of redundancy, resulting in lower performance, and in many cases intractability. Three application domains are presented: reasoning about feature-based variability of Software Product Lines (SPLs), probabilistic programming, and approximate programming.

\end{abstract}

\input{intro}
\input{motivation}
\input{vision}

\input{apps}
\input{conclusion}

\section* {Acknowledgments}
The author thanks the anonymous reviewers for their feedback and insightful suggestions.

\balance
\bibliographystyle{IEEEtran}
\bibliography{spl,datalog,logic,se}
\end{document}

%% file: macros.tex
\usepackage{caption}
\usepackage{subcaption}
\usepackage{proof}
\usepackage{algorithm2e}
\usepackage{color}
\usepackage{xcolor}
\usepackage{soul}

\newcommand{\term}[1]	{\emph{\textcolor{teal}{#1}}}
\newcommand{\crossed}[1] {\textcolor{red}{\st{#1}}}

\newcommand{\BD}{\begin{definition}}
\newcommand{\ED}{\end{definition}}

\newcommand{\code}[1]{\texttt{#1}}

%% file: macros_spl.tex
\newcommand{\lift}[1]{#1^\uparrow}

%% file: intro.tex
\section{Introduction}
\label{sec:intro}

Computer programs take concrete values as inputs, perform a sequence of operations on those inputs (assuming sequential, deterministic programs), and generate concrete outputs. All values involved in this process (inputs, intermediate values, and outputs) can be considered an abstraction of the \term{world} in which the program operates. A single world is typically assumed by a program. For example, a program variable \code{v} of type \code{T} maps to a single value of that type at any point in time, and a conditional expression takes a single branch based on what its guard evaluates to.

In the realm of mathematical logic, reasoning about multiple worlds at the same time has been the motivation behind \term{modal logics}. For example, temporal logics reason about not only the world at the current instance of time, but also about different instances (worlds) in the future. The \emph{S5} logic~\cite{Chellas:1980} is another example, distinguishing \emph{essential} truth (true in all worlds) and \emph{possible} truth (true only in some worlds). Modal logics have many applications in numerous domains, including verification of concurrent systems~\cite{Clarke:2018}, reasoning about computer programs~\cite{Pratt:1980}, artificial intelligence~\cite{McDermott:1980}, robotics~\cite{Fainekos:2005}, linguistics~\cite{Moss:2007}, and philosophy~\cite{Humberstone:2016}. This is a strong indication of how complex real-life problems are, and that new logical formalisms are continuously being introduced to address them effectively.

In several practical domains, software systems are also required to compute over inputs coming from multiple, potentially overlapping, worlds. Values varying from one world to another are appropriately labeled to ensure that outputs are also labeled accordingly. For example, a search engine storing multiple versions of a database has to effectively index and query multiple logical databases, each considered a world. Physically though, since those versions are only slightly different from one another, it is highly inefficient to store them and perform queries over them separately. Mining Software Repositories (MSR)~\cite{Kagdi:2007} is an example of a domain where such highly overlapping databases (successive versions of software artifacts) are common, and the scalability of different MSR systems is considered a challenge both to researchers and practitioners because of that~\cite{Shang:2010}.

In this paper we follow the analogy of modal logics, and argue that computing over multiple worlds is common around us, and needs to be addressed systematically (Sec.~\ref{sec:motivation}). The high-level logic of a program operating over multiple worlds is essentially the same as that of a single-world program. At the implementation level though, it needs to compute over multi-valued variables, where each value is appropriately labeled by an identifier of the world(s) it belongs to. The implementation also needs to be \emph{efficient}, i.e., with as few redundant computations as possible. We also argue that since the high-level program logic is oblivious to the number of worlds it is operating on, turning a single-world program into a modal (multi-world) one can be performed automatically, in a way similar to compiler optimizations. We provide a vision for rewriting single-world programs into their efficient modal counterparts in Sec.~\ref{sec:vision}.

\begin{figure}[t]
	\begin{subfigure}{0.5\textwidth}
		\includegraphics[width=\textwidth]{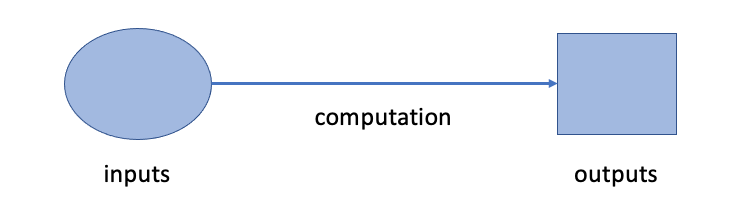}
		\caption{A computation maps inputs to outputs.}
		\label{fig:comp}
	\end{subfigure}
	
	\begin{subfigure}{0.5\textwidth}
		\includegraphics[width=\textwidth]{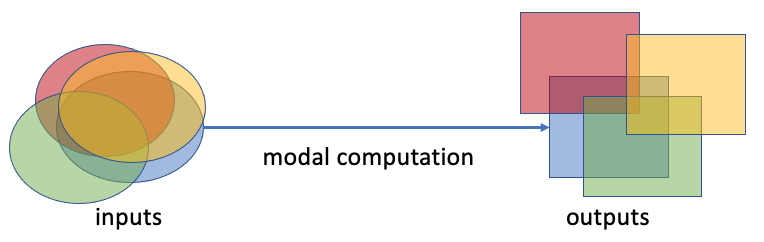}
		\caption{A modal computation maps inputs from multiple overlapping worlds, to potentially overlapping outputs corresponding to those worlds.}
		\label{fig:modalComp}
	\end{subfigure}
	\caption{Computation vs. modal computation.}
	\label{fig:comps}
	\vspace{-0.2in}
\end{figure}

Fig.~\ref{fig:comps} illustrates the difference between single-world computations (Fig.~\ref{fig:comp}), mapping a single set of inputs to a single set of outputs, and modal computations (Fig.~\ref{fig:modalComp}), working on multiple potentially overlapping sets of inputs at the same time and generating outputs for each world, also potentially overlapping. There are cases where each world is explicitly labeled, and others in which they can be only quantitatively described. Examples of each are discussed in Sec.~\ref{sec:apps}.

%% file: motivation.tex
\newcommand{\FA}{FA}
\newcommand{\FB}{FB}

\section{Motivation}
\label{sec:motivation}

This work has been originally motivated by the analysis of \term{Software Product Lines (SPLs)}. An SPL is a family of related software products, developed together from a common set of artifacts~\cite{Clements:2001}. The unit of variability in an SPL is a \term{feature}. Each feature can be either present or absent in each of the product variants of the SPL. Given this combinatorial nature of features, the number of products grows exponentially with the number of features. Each product is defined by a \term{feature configuration}, which is the set of features present in that product. A \term{feature expression} is a propositional formula over features, denoting a set of products.

Efficiently analyzing SPLs has been an active area of software engineering and programming languages research for more than a decade. Software analyses (of source code, transition system models, graphical models, etc...) typically work on a single product at a time, and they need to be modified to be applicable to a whole SPL. This is an instance of the problem of taking a single-world program, and turning it into a semantically equivalent multi-world one.

The product line example in Listing~\ref{lst:spl} has two features: \FA~and \FB. Assuming all product combinations are valid, four products can be generated from this example with the following feature combinations: \{\}, \{\FA\}, \{\FB\}, and \{\FA, \FB\}. Assume we have a static program analysis for detecting division-by-zero errors. Only one of the four configurations (\{\FB\}) has a statically detectable division-by-zero error, where the value of variable \code{c} is not incremented in line 5 because \FA~is absent, and line 9 (specific to feature \FB) divides \code{(x + y)} by \code{c}. The original single-world analysis can be applied to each of the four products of this SPL individually, reporting results for each. However, real-life SPLs usually have hundreds, sometimes thousands, of features (The Linux kernel has more than 10,000 features~\cite{Nadi:2014}), so enumerating all products and analyzing them one-by-one is not generally feasible. In addition, different product variants have a lot in common, and leveraging this amount of commonality can make analyzing the whole product line at once much more efficient than analyzing the products one at a time.

\begin{figure}
\begin{lstlisting}[language=c,numbers=left,morekeywords=ifdef endif,xleftmargin=15pt]
int foo(int x, int y)
{
    int c = 0;
#ifdef FA
    c++;
#endif

#ifdef FB
    return (x + y) / c;
#else
    return (x + c) / y;
#endif
}
\end{lstlisting}

\captionof{lstlisting}{Example of a Software Product Line (SPL).}
\label{lst:spl}
\end{figure}

Several analyses of different kinds have been manually redesigned and re-implemented to support variability~\cite{Kastner:2011, Gazzillo:2012, Classen:2013, Thum:2014, Salay:2014, Rhein:2018}. Re-implementing an analysis is a lengthy and error-prone process, so several attempts have been made to come up with systematic or automated approaches to this process. Examples include systematic variability-aware abstract interpretation~\cite{Midtgaard:2015}, SPL data-flow analyses based on the IFDS framework~\cite{Reps:1995, Bodden:2013}, variability-aware analyses written in Datalog~\cite{Shahin:2019}, and program analyses written in a functional language~\cite{Shahin:2020b}.

We argue that the research effort that has been put in SPL analysis (also known as \term{variability-aware lifting}) can be generalized to domains other than SPLs, where efficient computations are needed over multiple overlapping worlds. In SPL analysis, each product is considered a world, labeled by a feature configuration. Inputs, outputs and intermediate values of the analysis are labeled by propositional formulas over features (known as \term{Presence Conditions (PCs)}) denoting sets of products. Other problem domains can be mapped into this same formulation of worlds. 

There are also cases where the individual worlds are not explicit, but rather quantified. For example, in a \term{probabilistic program}, a variable \code{v} can have the value \code{\{(7,0.2), (9,0.8)\}}, i.e., the value \code{7} with probability 0.2 (20\% of possible worlds), and the value 9 with probability 0.8 (80\% of possible worlds). Another example is \term{approximate computing}, where instead of a concrete value, a variable can at any point in time be defined by a minimum-maximum interval. For example, \code{v = [4..9]} means v is approximated to the range 4 to 9 inclusively.

%% file: vision.tex
\newcommand{\lit}[1]  {\code{#1}}
\newcommand{\x}{\code{x}}
\newcommand{\y}{\code{y}}
\newcommand{\z}{\code{z}}

\section{Vision}
\label{sec:vision}

\begin{figure*}[t]
	\begin{tabular}{lc}
		\begin{tabular}{l}
			\begin{subfigure}[]{0.59\textwidth}
				
				\code{x} = \{(\lit{-7},$\FA$),(\lit{3}, $\neg \FA$)\}
				
				\code{y} = \{(\lit{1}, $\FA \wedge \FB$), (\lit{8},$\FA \wedge \neg \FB$), (\lit{4}, $\neg \FA \wedge \FB$), (\lit{10}, $\neg \FA \wedge \neg \FB$)\}
				
				\code{z} = \{(\lit{5}, $\texttt{True}$)\}
				
				\caption{Arguments \code{x}, \code{y}, and \code{z}.}
				\label{fig:args}
			\end{subfigure}
		    \\
		    \begin{subfigure}[]{0.59\textwidth}
		    	
		    	\begin{lstlisting}
		    	
int foo(int x, int y, int z) {
    return bar(x, y) + baz(z);
}
		    	\end{lstlisting}
		    	\caption{Function \code{foo}.}
		    	\label{fig:foo}
		    \end{subfigure}
		    \vspace{0.1in}

			\vspace{0.1in}
			\\
				

\end{tabular}
&
\begin{subfigure}[]{0.3\textwidth}
\begin{tabular}{cccl}
\hline
$\x$ & $\y$ & $\z$ & \textit{\textbf{LABEL}} \\
\hline
-7 & 1 & 5 & $\FA \wedge \FB$ \\
-7 & 8 & 5 & $\FA \wedge \neg \FB$ \\
\crossed{-7} & \crossed{4} & \crossed{5} & \crossed{$\FA \wedge \neg \FA$} \\
\crossed{-7} & \crossed{10} & \crossed{5} & \crossed{$\FA \wedge \neg \FA$} \\
\crossed{3} & \crossed{1} & \crossed{5} & \crossed{$\neg \FA \wedge \FA$} \\
\crossed{3} & \crossed{8} & \crossed{5} & \crossed{$\neg \FA \wedge \FA$} \\
3 & 4 & 5 & $\neg \FA \wedge \FB$ \\
3 & 10 & 5 & $\neg \FA \wedge \neg \FB$ \\
\hline
\end{tabular}
\caption{Input vectors for $\code{foo}$.}
\label{fig:inputs}
\end{subfigure}
\end{tabular}
\caption{Variability-aware application of function \code{foo} to modal arguments \code{x}, \code{y}, and \code{z}, tracing the cross-product of modal arguments (adapted from~\cite{Shahin:2020b}).}
    \vspace{-0.1in}
\label{fig:fooExample}
\end{figure*}

Mathematical logic provides a formal framework for reasoning about truth. There are many cases though where truth and falsehood are not absolute; for example a statement can be true now but not in the future, or it can be true from one person's perspective and false from another's. Modal logics~\cite{Blackburn:2001, Huth:2004} add qualifiers to logical formulas, and those qualifiers denote modalities, or modes of truth. For example, in the modal logic of necessity and possibility \emph{S5}~\cite{Chellas:1980}, the $\Box$ connective indicates necessity (truth in all possible worlds), while the $\Diamond$ connective indicates possibility (truth in some possible world(s)). Those are two unary connectives that qualify the truth of propositional formulas. 

Given how logic and computation are sometimes viewed as two sides of one coin (e.g., the Curry-Howard isomorphism~\cite{Nederpelt:2014}), computing, rather than just reasoning, about multiple worlds also has potentially many applications. Examples include explicit variability in Software Product Lines (SPLs), and quantitative variability in probabilistic models. 

Our goal is to take a single-world program, and automatically convert it into a semantically equivalent multiple-world (or modal) program.  Modalities labeling the individual worlds or quantifying them, and the semantics of those modalities, are orthogonal to the rewriting process. We refer to the process of modal program rewriting as \term{modal lifting}.

For example, function \code{foo} in Fig.~\ref{fig:foo} takes three parameters of type \code{int}. A semantically equivalent lifted function would take three arguments of type $\lift{\code{int}}$ instead, where values of type $\lift{\code{int}}$ are sets of \code{(int, LABEL)} pairs. Variables \x, \y, and \z~in Fig.~\ref{fig:args} are of type $\lift{\code{int}}$, where labels are propositional formulas identifying sets of products in a software product line with two features \FA~and \FB.

Each label within a modal value denotes a set of worlds, and has to satisfy two conditions:
(1) \term{Disjointness}: within the same modal value, the sets of worlds denoted by the different labels have to be mutually disjoint. If they are not, then we would be allowing different values within the same world, which would result in non-deterministic evaluation of programs.
(2) \term{Totality}: within the same modal value, the union of all the sets of worlds denoted by the different labels has to be equal to the set of all worlds. This ensures that each modal value is defined in each world.

In the case of software product lines, disjointness and totality on feature expressions are defined in~\cite{Shahin:2020b}.
Disjointness is defined as the propositional unsatisfiability of the conjunction of any pair of different labels within the same modal value. For example, the variable \y~in Fig.~\ref{fig:args} has four labels: $\FA \wedge \FB$, $\FA \wedge \neg \FB$, $\neg \FA \wedge \FB$, and $\neg \FA \wedge \neg \FB$. Conjoining any two of them results in an unsatisfiable propositional formula. Totality on the other hand is defined as the disjunction of labels within a modal value resulting in a tautology. Again, this is the case for each of the variables in Fig.~\ref{fig:args}.

In the context of software product lines, two automatic program lifting approaches have been proposed in~\cite{Shahin:2020b}: \term{shallow lifting} and \term{deep lifting}. Shallow lifting takes a program as a black-box and generates the cross-product of program arguments, passing each combination of arguments to the original single-world program. For example, tuples of the cross-product of arguments in Fig.~\ref{fig:args} are listed in Fig.~\ref{fig:inputs}.

Algorithm~\ref{alg:shallow} outlines how an n-ary function $f$ is shallow lifted and applied to modal arguments $arg_1,...,arg_n$. First, we calculate the cross product of the arguments (each of which is a set of pairs). Each element in the cross-product is an n-tuple of $(v,label)$ pairs. Within each element, if the intersection of the sets of worlds denoted by the labels is the empty set, then we can safely ignore this element. However, if the intersection is not empty, we apply the function $f$ to the arguments $v_1,...,v_n$. and add the result to the output set, which is returned at the end.

\begin{algorithm}
	\SetAlgoLined
	\KwIn{$f$, $arg_1$, ..., $arg_n$}
	\KwResult{applying shallow lifting of $f$ to $args$}
		$output = \{\};$ \\
		\ForEach
		{ $((v_1,label_1),...,(v_n,label_n)) \in (arg_1 \times ... \times arg_n)$ }
		{
			\If{($\bigcap_{i=1}^{n} label_i \neq \phi$) }
			{ $output = output \cup \{f(v_1,...,v_n)\}$; }
		}
\KwRet $output$;
\vspace{0.1in}
\caption{Shallow lifting algorithm.}
\label{alg:shallow}
\end{algorithm}
\vspace{-0.1in}

\newcommand{\unsat}{\code{unsat}}

\begin{table*}[t]
	\centering
	\begin{tabular}{lccc}
		\hline
		& Software Product Lines & Probabilistic Programming & Approximate Programming \\
		\hline
		Label Representation & Propositional formula & Probability $\in [0..1]$ & MIN and MAX \\
		Cardinality & > 0 & > 0 & 2 \\
		Intersection & conjunction & multiplication & Group by MIN, MAX\\
		Union & disjunction & addition & Group by MIN, MAX\\
		Emptiness & unsatisfiability & 0.0 & $v_{MAX} < v_{MIN}$ \\
		Disjointness & $\forall i \neq j \cdot \unsat (l_i \wedge l_j)$ & $\Sigma_{i} {l_i} \leq 1.0$ & 1 MIN and 1 MAX  \\
		Totality & $\bigvee_{i} {l_i} = \code{True}$ & $\Sigma_{i} {l_i} = 1.0$ & 1 MIN and 1 MAX \\
		\hline
	\end{tabular}
	\caption{Comparing modalities, their invariants, and operations across three application domains: feature-based reasoning of software product lines, probabilistic programming, and approximate programming.}
	\label{tbl:modalities}
\end{table*}

Applying the shallow lifting algorithm to the example in Fig.~\ref{fig:fooExample}, the first step is calculating the cross-product of arguments \code{x}, \code{y}, and \code{z} (Fig.~\ref{fig:inputs}). For each row in the table, \emph{LABEL} is the conjunction of presence conditions of the components of \code{x}, \code{y}, and \code{z} in this row. Recall that conjunction is the logical operator corresponding to intersection of sets denoted by propositional formulas (feature expressions). Rows crossed out have unsatisfiable labels, i.e., they correspond to the empty set of worlds, so they are ignored, leaving out four input vectors.

Shallow lifting suffers from the inability to exploit commonalities and eliminate redundancies, and thus can be further optimized. For example, function \code{foo} in Fig.~\ref{fig:foo} internally passes \code{x} and \code{y} to \code{bar}, and passes \code{z} to \code{baz}. Assuming the values in Fig.~\ref{fig:args}, \code{z} happens to be the constant value \code{5} across all configurations, so shallow-lifted \code{foo} ends up calling \code{baz} four times, passing the same argument each time. 
Shallow lifting treats \code{foo} as a black-box, so the opportunity to eliminate the redundant calls is not visible. 

\newcommand{\intUp}{$\lift{\code{int}}$}
\newcommand{\fooUp}{$\lift{\code{foo}}$}

Deep lifting on the other hand inspects the internals of a program, rewriting it to maximize sharing of common values, and to minimize redundancy. The idea behind deep lifting is to statically push the calculation of cross-products of modal values as deep as possible down the call tree of a program. For example, \code{foo} is re-written into $\lift{\code{foo}}$ as follows:

\begin{lstlisting}[escapeinside={(*}{*)}]

(*\intUp*) (*\fooUp*) ((*\intUp*) x, (*\intUp*) y, (*\intUp*) z) {
   return 
      shallowLift((+),
                  shallowLift(bar, x, y),
                  shallowLift(baz, z));
}
\end{lstlisting}

Function \code{foo} internally calls three functions/operators: \code{bar}, \code{baz}, and the (+) operator. Each of the three can be shallow lifted using Algorithm~\ref{alg:shallow}, and their shallow lifted counterparts operate on \intUp~rather than \code{int} arguments. This way, the expression \code{shallowLift(baz, z)} within \fooUp~calls the underlying function \code{baz} on the single component of argument \code{z} only once. If any of \code{bar} or \code{baz} is not a primitive function, it can be similarly rewritten into a deep-lifted function, which is then called from within \code{foo} instead of using \code{shallowLift}. This process can recursively applied down the call tree, rewriting the whole program. Deep lifting rewrite rules for a subset of Haskell including conditional expressions, pattern matching, and polymorphic lists have been presented in~\cite{Shahin:2020b} in the context of lifting program analyses to software product lines.


%% file: apps.tex
\section{Applications}
\label{sec:apps}

In this section we show how modal rewriting is oblivious to the specific modality used, which means the same lifting approach can be used across different kinds of modalities. We compare three application domains: Featured-based variability of software product lines, probabilistic programming, and approximate programming. The modalities of each application domain, their invariants and operators are summarized in Table~\ref{tbl:modalities}.

\subsection{Feature-based Variability}

When reasoning about Software Product Lines (SPLs), each feature is represented as a propositional symbol, and a set of products is denoted by propositional formulas over features (usually referred to as \term{feature expressions}). The set of worlds in this context is the set of products, and labels within modal values are feature expressions.

A modal feature-based value has at least one element (cardinality > 0). Intersection of sets of products denoted by feature expressions is conjunction of feature expressions. Similarly, union is disjunction. A set denoted by a feature expression is empty when the feature expression is an unsatisfiable formula.

Disjointness of feature expressions is the same as set disjointness: sets are disjoint when their pair-wise intersection is empty. Translating this to feature expressions, a collection of feature expressions is disjoint if their pair-wise conjunction is unsatisfiable. Similarly, totality of a collection of sets is the universe being equal to the union of that collection. In propositional logic, the tautology \code{True} is analogous to the universe of all worlds, and disjunction is analogous to set union.
 
\subsection{Probabilistic Programming }

Probabilistic programming surprisingly has a lot in common with feature-based variability. The main difference is that feature expressions define explicit sets, while probabilities quantify over them. For example, a probability of 0.8 indicates 80\% of all worlds, without naming which 80\%. This kind of quantitative description of sets has many applications, particularly in modeling and simulation.

Labels for probabilistic modal values come from the range [0.0..1.0]. Cardinality has to be greater than zero, same as in feature-based modalities. Assuming probabilistic independence, intersection of two sets denoted by probabilities is the multiplication of those probabilities. Similarly, union is their addition, and a probability of 0.0 denotes the empty set. 

Because probabilities do not define explicit sets, disjointness is also quantitative rather than explicit. A necessary condition (but not always adequate) for disjointness of a collection of sets denoted by probabilities is that the sum of those probabilities is at most 1.0. Totality makes this bound more strict though, turning the inequality into an equality.

\subsection{Approximate Programming}

Approximate programming has many applications, particularly in software systems running in power-saving mode~\cite{Mittal:2016}, and systems reading inputs from imprecise sensors. Approximation modalities can be represented in many ways, including a fidelity of a single value, or a range of possible values (minimum and maximum) over a continuous domain. We follow the latter representation in this example.

Cardinality here is fixed at two, since each modal range value will have exactly two sub-values: one labeled as MIN and the other labeled as MAX. Values in this MIN..MAX range form a uniform distribution of possible values. Intersection and union are again different in this case because they both boil down to grouping MIN values together and MAX values together. A range is empty when the value labeled by MAX is less than that labeled by MIN.

Similarly, the disjointness and totality invariants are enforced by the fact that there is exactly one MIN value and exactly one MAX value. Lifted programs will compute results for the MIN value and MAX value, labeling the results. accordingly. 

%% file: conclusion.tex
\section{Conclusion and Future Work}
\label{sec:conclusion}

In this paper we introduced the notion of Modal Software Engineering: automatically turning sequential, deterministic programs into semantically equivalent programs, that efficiently operate on inputs coming from multiple intersecting worlds. We drew an analogy between modal logics, and software application domains where multiple sets of potentially overlapping inputs (multiple worlds) need to be processed efficiently. Processing each world independently instead would result in a lot of redundancy, resulting in lower performance, and in many cases intractability. Three application domains have been presented: reasoning about feature-based variability of Software Product Lines (SPLs), probabilistic programming, and approximate programming.

For future work, we plan to investigate the three example domains presented in this paper in more depth, identifying domain-specific problems and reflecting on whether they are related to problems in other domains. Examples of such problems are dependent variables in probabilistic programs, and approximate computing over categorical values. An orthogonal research direction we plan to pursue is studying the different program rewriting techniques (and their trade-offs) for lifting programs into modal ones.